\documentstyle[epsf,11pt]{article}

\newcommand{\be}{\begin{equation}}
\newcommand{\ee}{\end{equation}}
\newcommand{\manifold}{F}

\title{Understanding multifractality: reconstructing images from edges}

\author{Antonio Turiel\thanks{e-mail: amturiel@delta.ft.uam.es}
 ~and Angela del Pozo\thanks{e-mail: angela@ipsa.es}\\
Computational Neuroscience Group\\
Departamento de F\'{\i}sica Te\'orica\\
Universidad Aut\'onoma de Madrid\\
Cantoblanco 28049, Madrid. SPAIN}

\begin{document}

\maketitle

\begin{abstract}
It has been recently proven \cite{PRLnuestro,Singularities} that natural
images exhibit scaling properties analogue to those of turbulent flows. 
These properties allow regarding each image as a \emph{multifractal} 
object \cite{Singularities}, for which its most singular manifold conveys 
the most of the non-redundant structure. In the present work, we go 
further in this analysis, proposing a simple propagator that 
reconstructs the whole image from this set. This fact could 
have deep implications for biology, technology and statistical mechanics.
\end{abstract}

\vspace{1cm}
\noindent
{\it Short title:} Reconstructing Lena

\noindent
Preprint number: FTUAM 98-23

\noindent
PACS numbers: 42.30.Wb, 47.53.+n, 47.54.+r, 87.10.+e

\indent
Our main motivation was the understanding of the coding strategies 
\cite{laughlin} developed by biological neural systems, which has been a 
matter of great interest, specially in the case of the visual pathway 
\cite{hateren,AR92}. This problem has often been analyzed using the 
Information Theory background\cite{Shannon,Cover}, thus introducing a 
statistical treatment of the visual signal \cite{Ruderman} . It is 
precisely a statistical analysis  based on a very different context (the 
Fully Developed Turbulence) which leads to the most interesting results.
This scope provides what seems a powerful new language; new for image
analyists but well known for physicist in the field of turbulence.
The structure of the paper is as follows: the theoretical basis about the
turbulent-like statistics of natural images is presented. This short 
review introduces the concept of Most Singular Manifold as a statistical
essential in images. Then, a guess about how could this set be used to
retrieve the whole image is issued and a simple interpretation of it is
given. This guess is self-consistently used to check that multifractality
is a robust, intrinsic property of natural images; and a possible
interpretation of it, perhaps valid beyond this subject, is proposed.

\indent
The work was developed on B/W images, although generalizations to coloured
images could be done in the same way. We consider each image to be defined
by the scalar field of its luminosities, $I(\vec{x})$, although we will
make use of the contrast $c(\vec{x})\equiv I(\vec{x})-\langle I(\vec{x})
\rangle_{\vec{x}}$. The main result of \cite{PRLnuestro} was to prove that
that it is possible design intensive variables, related to the derivatives
of $c(\vec{x})$, which exhibit scaling laws known as Self Similarity (SS)
and Extended Self Similarity (ESS). These properties are the same as
those of turbulent flows, which allowed for using models developed in such
a context. It is also possible reinterpret that result in terms of a
{\it multifractal measure} \cite{PF85,Arneodo}. Very recently
\cite{Singularities} has been established that for natural images
the scalar density $|\nabla c|^2(\vec{x})d\vec{x}$ defines a multifractal
measure, i.e.: The image may be decomposed as a collection of fractal sets
$\manifold_h$ (each having fractal dimension $D(h)$), formed by the points
for which the measure scales locally as $r^h$. This characterization is
observed indirectly by the scaling of $p$-moments related to the measure,
which in turn shows the SS properties (as it is comprehensively described
in \cite{Singularities}).

\begin{figure}[htb]
\begin{center}
\hspace*{0cm}
\epsfxsize=8cm
\epsffile{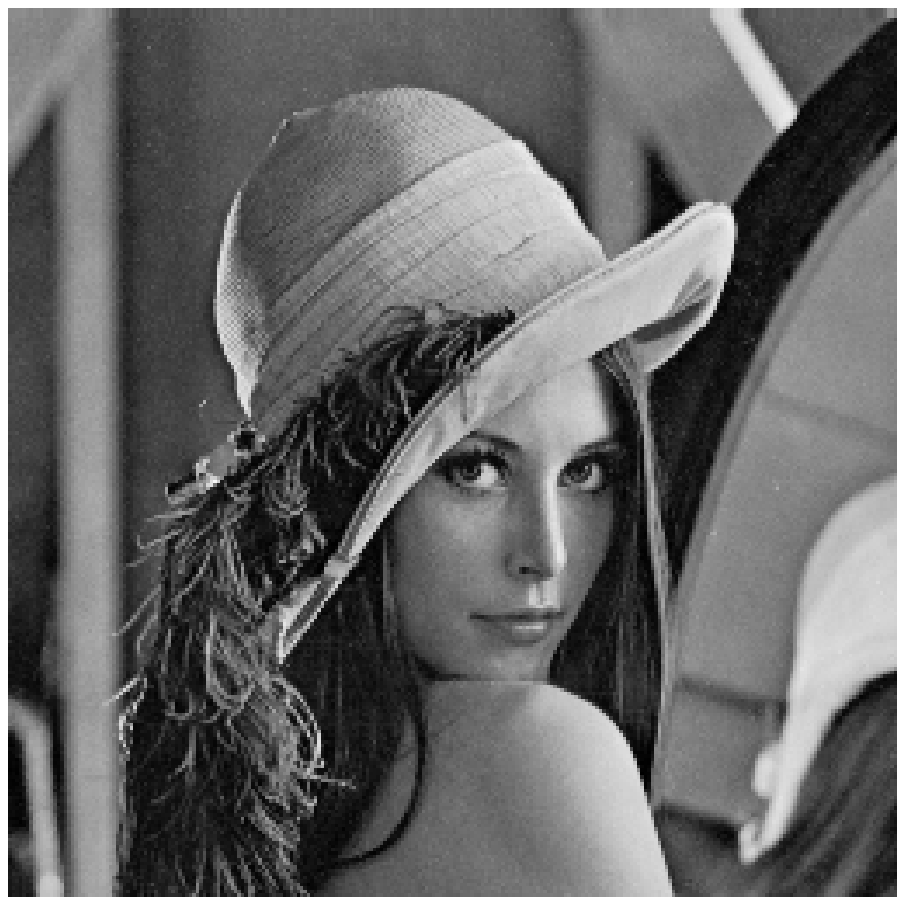}
\end{center}
\caption{Lena's image}
\label{fig:Lena}
\end{figure}

\begin{figure}[htb]
\hspace*{0cm}
\begin{center}
\epsfxsize=8cm
\epsffile{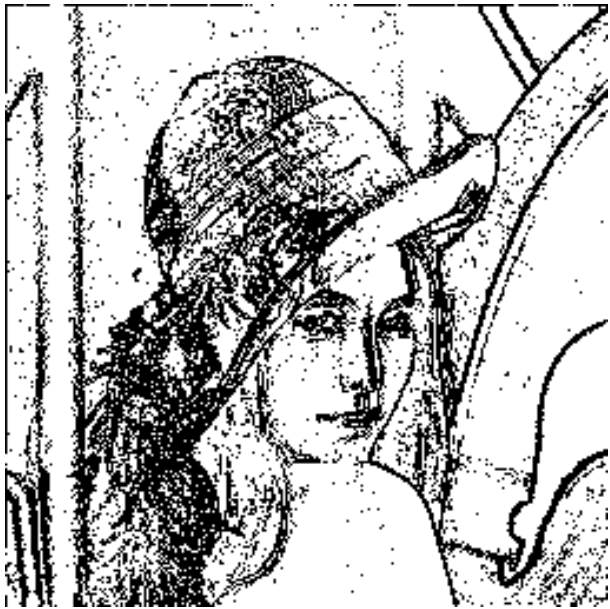}
\end{center}
\caption{Lena's most singular manifold}
\label{fig:Lena_MSM}
\end{figure}

\indent
Our first step was to split an image into its different fractal sets.
This can easily be done by means of the wavelet analysis (see
\cite{Daubechies}) , as a multiresolution decompositon (see 
\cite{Arneodo,Mallat}). We followed the detailed technique presented in
\cite{Singularities}. One also needs to know the characteristic values of
the multifractal (namely, the exponent associated to the Most Singular
Manifold (MSM) ($h_{\infty}=-0.5$) and the fractal dimension of this set
($D_{\infty}=1.$; see \cite{PRLnuestro,Singularities})). 
The analysis revealed that the MSM,
$\manifold_{\infty}$, can be identified with what could naively be called
``edges'' of the objects present in the scene (see Figures \ref{fig:Lena} 
and \ref{fig:Lena_MSM}). More interestingly, the multiplicative process
model \cite{Novikov,Castaing} used to describe multifractal structures
\cite{PF85,SheLeveque} suggests that there is a hierarchical organization
of the different fractal manifolds in the image, as it is shown in
\cite{Singularities}. The open question is if one of them (namely the
MSM) contains enough information to deterministically reconstruct
the whole set. This is precisely the aim of this letter.

\indent
We consider the gradient $\nabla c(\vec{x})$ as our basic field. We would
like to reconstruct the field $c(\vec{x})$ for every $\vec{x}$ of the
scene given the gradient on $\manifold_{\infty}$. Under the assumptions of
translational invariance and linearity, one obtains:

\be
c(\vec{x}) = \int_{\manifold_{\infty}} dl(\vec{y})\;
\vec{g}(\vec{x}-\vec{y}) \nabla c(\vec{y})
\label{kernel}
\ee

\noindent
where $\vec{g}$ is the linear kernel of the desired propagator, and
$\int_{\manifold_{\infty}} dl(\vec{y})$ means line integration along the
MSM. This equation can be rewritten in a very useful
form defining the field $\vec{v}_0$ as

\be
\vec{v}_0(\vec{x}) =\nabla c (\vec{x})
\delta_{\manifold_{\infty}}(\vec{x})
\ee

\noindent
where $\delta_{A}(\vec{x})$ stands for the proper Hausdorff measure
especialized to the set A. In this way, eq.~(\ref{kernel}) is elegantly
expressed in the Fourier space as

\be
\hat{c}(\vec{f}) = \hat{\vec{g}}(\vec{f}) \cdot \hat{\vec{v}}_0 (\vec{f})
\label{Fourierkernel}
\ee

\noindent
which is an integral equation equivalent to eq.~(\ref{kernel}), but now 
the boundary conditions are contained in the vector field
$\vec{v}_0$, which depends on the particular image to be reconstructed.
The crucial point in all that follows is to determine $\hat{\vec{g}}$.
It is natural to require it to be isotropic, as the particularities of the
image are already contained in $\vec{v}_0$, and we think that $\hat{\vec{g}}$
is an universal propagator. This would imply
$\hat{\vec{g}}(\vec{f})\propto
\vec{f}$. To end with, we recall a well established property of 
natural images, namely the scaling of their power spectrum $S(\vec{f})$
(see \cite{Field}), which is:

\be
S(\vec{f})\equiv |\hat{c}(\vec{f})|^2 \sim \frac{1}{f^{2-\eta}},
\label{power}
\ee

\noindent
where $\eta$ is a non-universal, small exponent which depends on the
particular image ensemble considered (see for instance \cite{Tollhurst}).
The simplest possible $\hat{\vec{g}}$ is then given by:

\be
\hat{\vec{g}}(\vec{f}) = \frac{i \vec{f}}{f^2}
\label{guess}
\ee

\begin{figure}[htb]
\begin{center}
\hspace*{0cm}
\epsfxsize=8cm
\epsffile{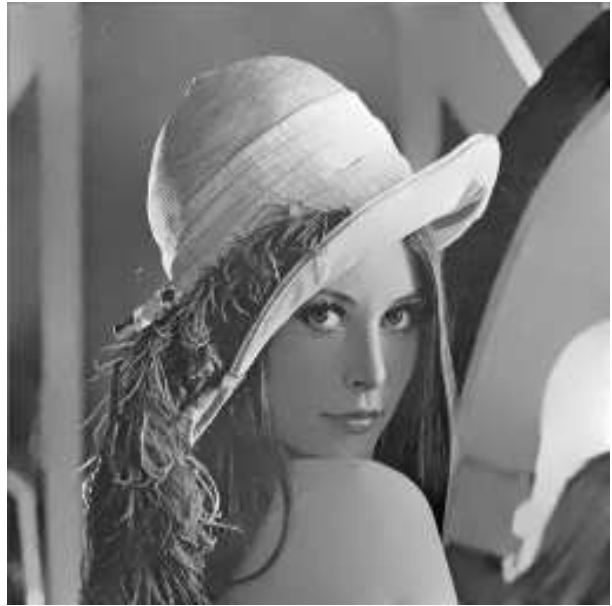}
\end{center}
\caption{Lena's reconstruction from her most singular manifold}
\label{fig:Lena_kernel}
\end{figure}

\begin{figure}[htb]
\vspace*{-2cm}
\begin{center}
\hspace*{-1cm}
\leavevmode
\epsfxsize=5.5cm
\epsffile{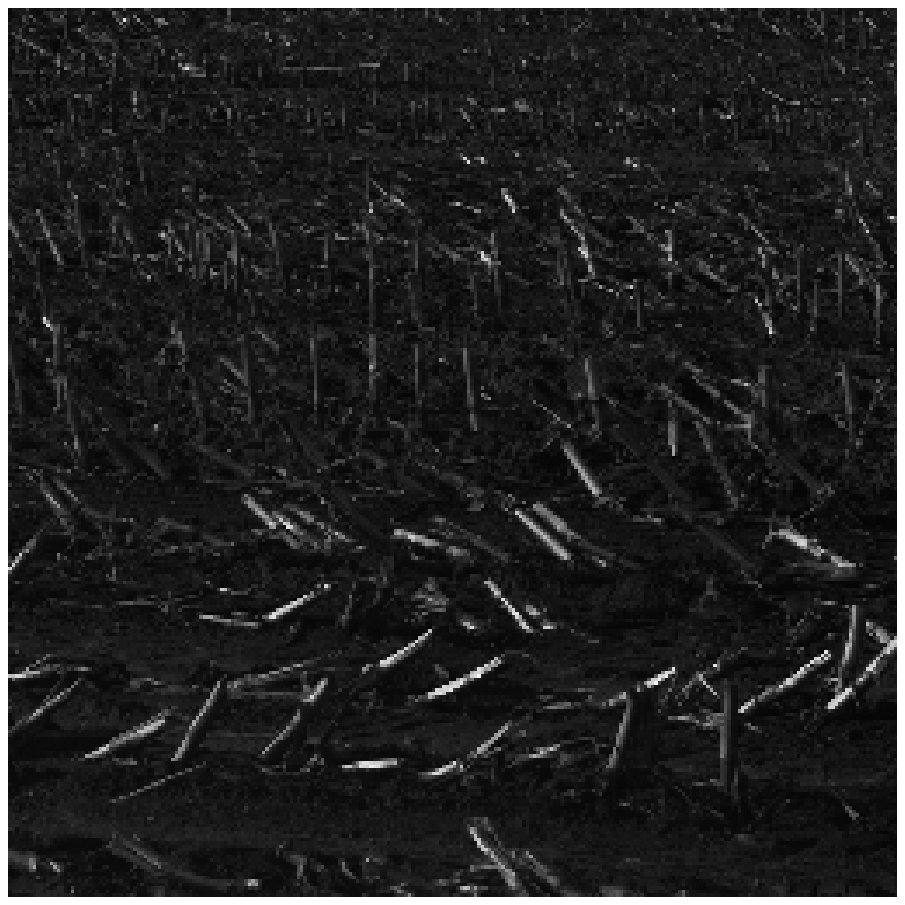}
\epsfxsize=5.5cm
\epsffile{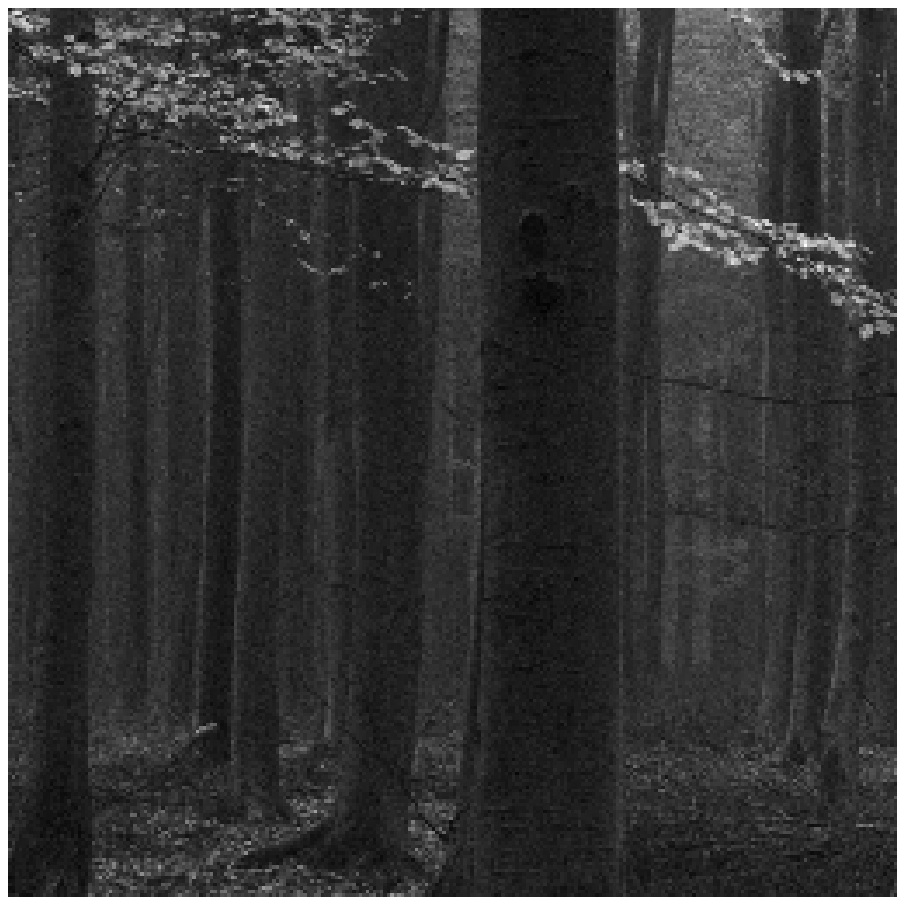}
\epsfxsize=5.5cm
\epsffile{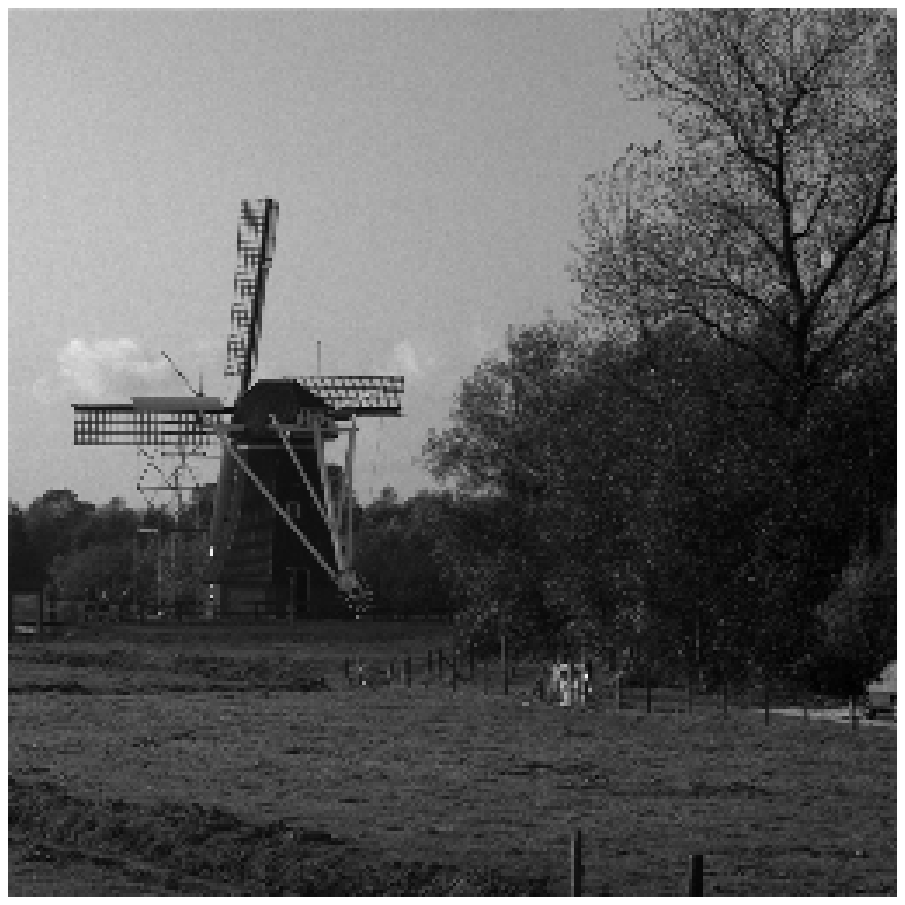}
\\
\vspace*{0.1cm}
\hspace*{-1cm}
\leavevmode
\epsfxsize=5.5cm
\epsffile{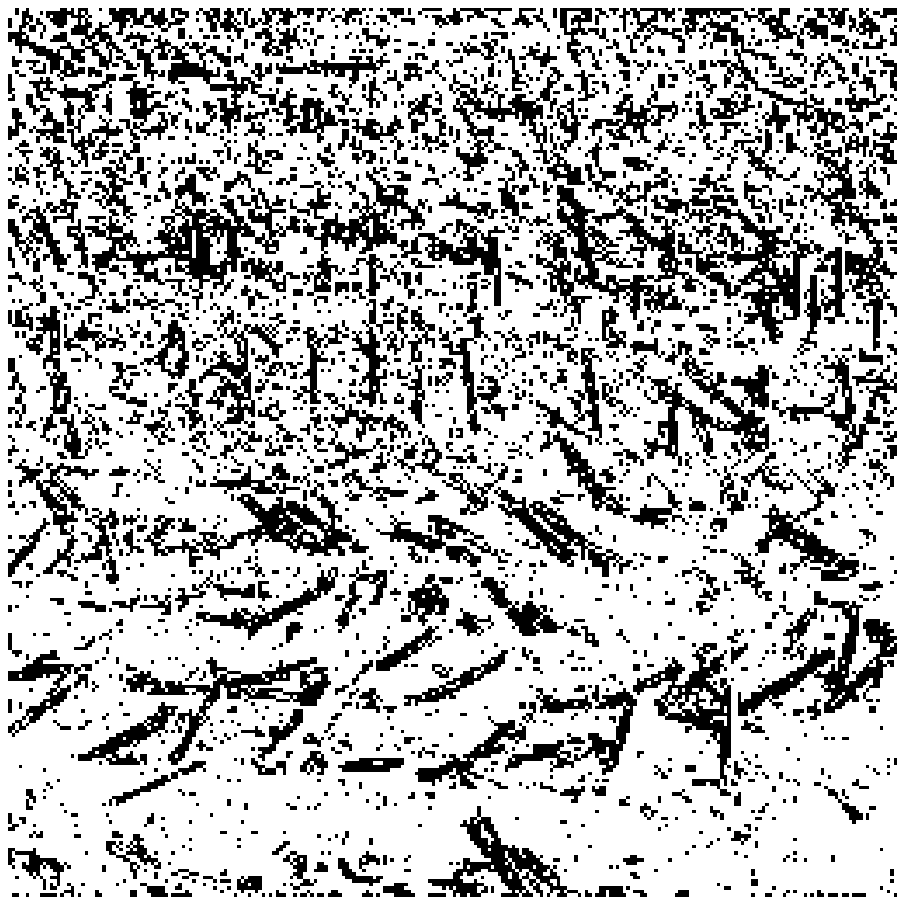}
\epsfxsize=5.5cm
\epsffile{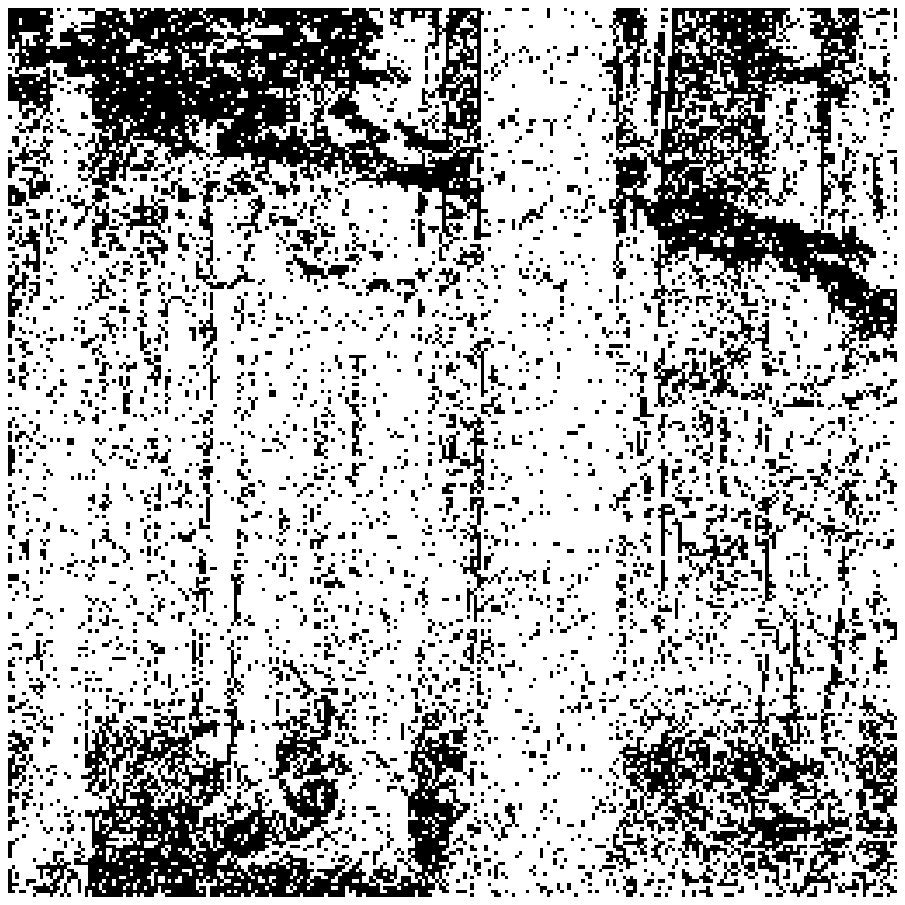}
\epsfxsize=5.5cm
\epsffile{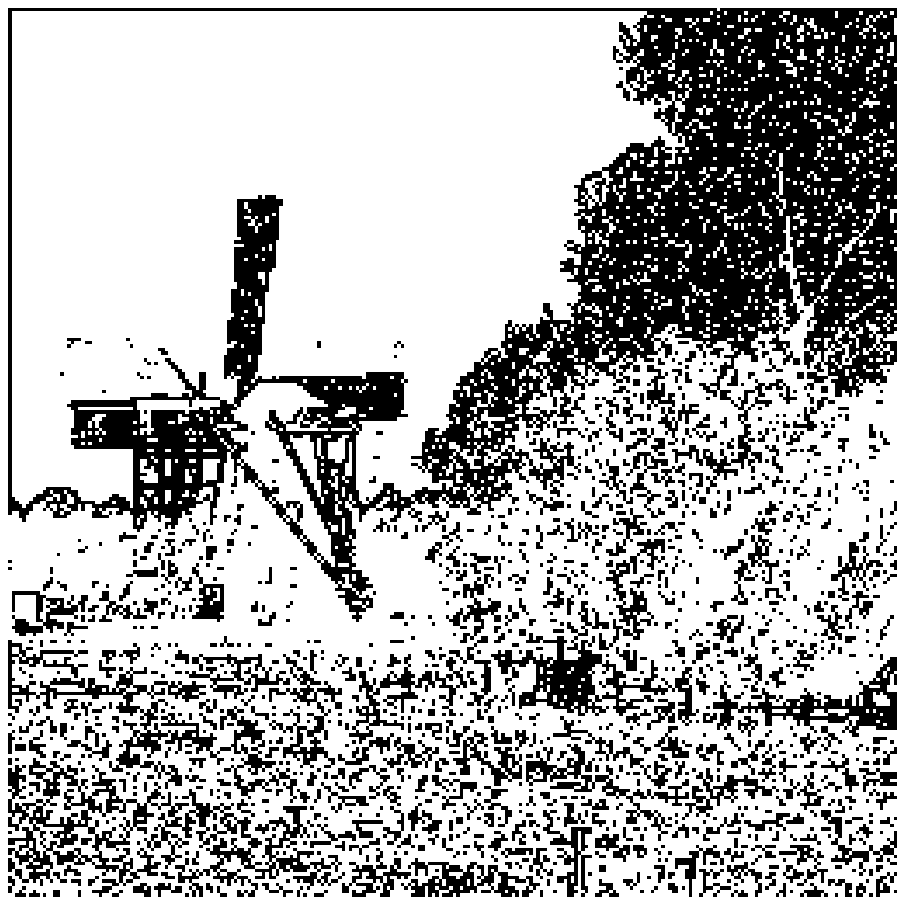}
\\
\vspace*{0.1cm}
\hspace*{-1cm}
\leavevmode
\epsfxsize=5.5cm
\epsffile{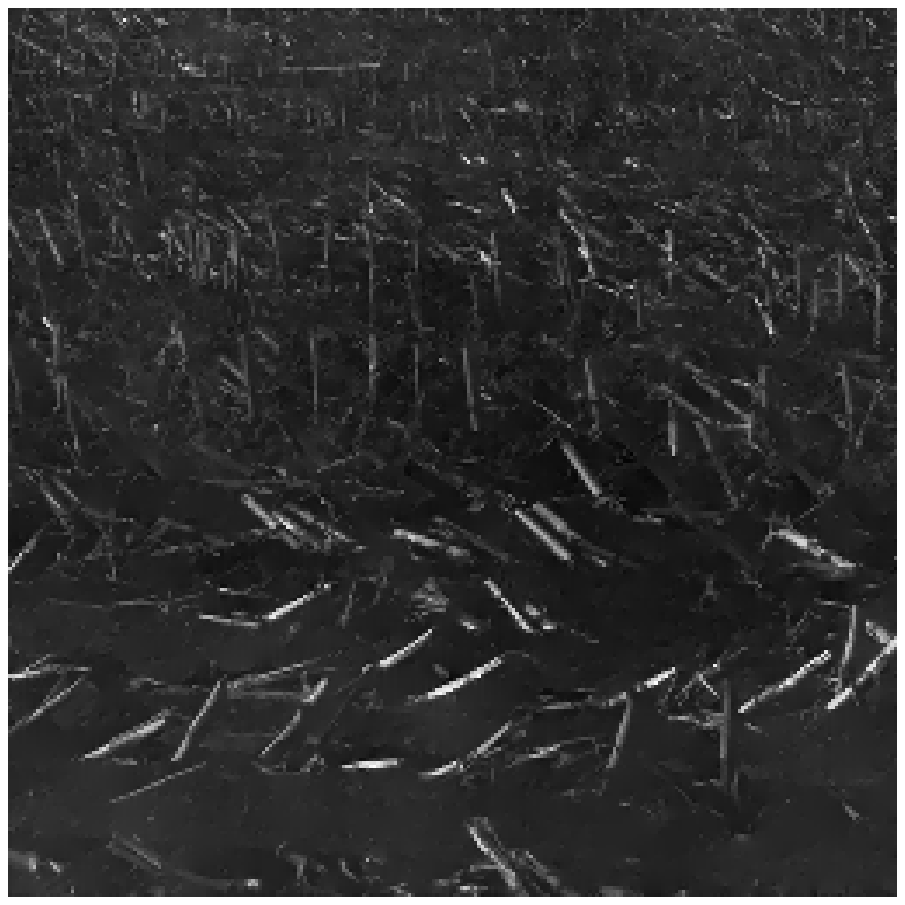}
\epsfxsize=5.5cm
\epsffile{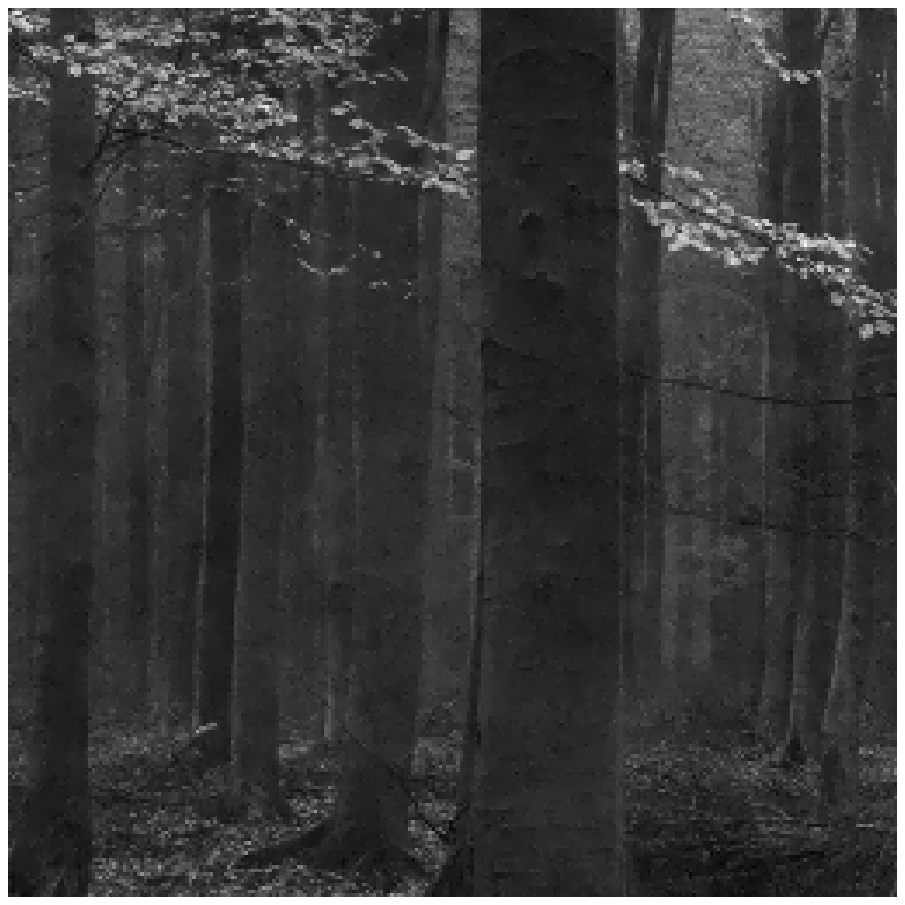}
\epsfxsize=5.5cm
\epsffile{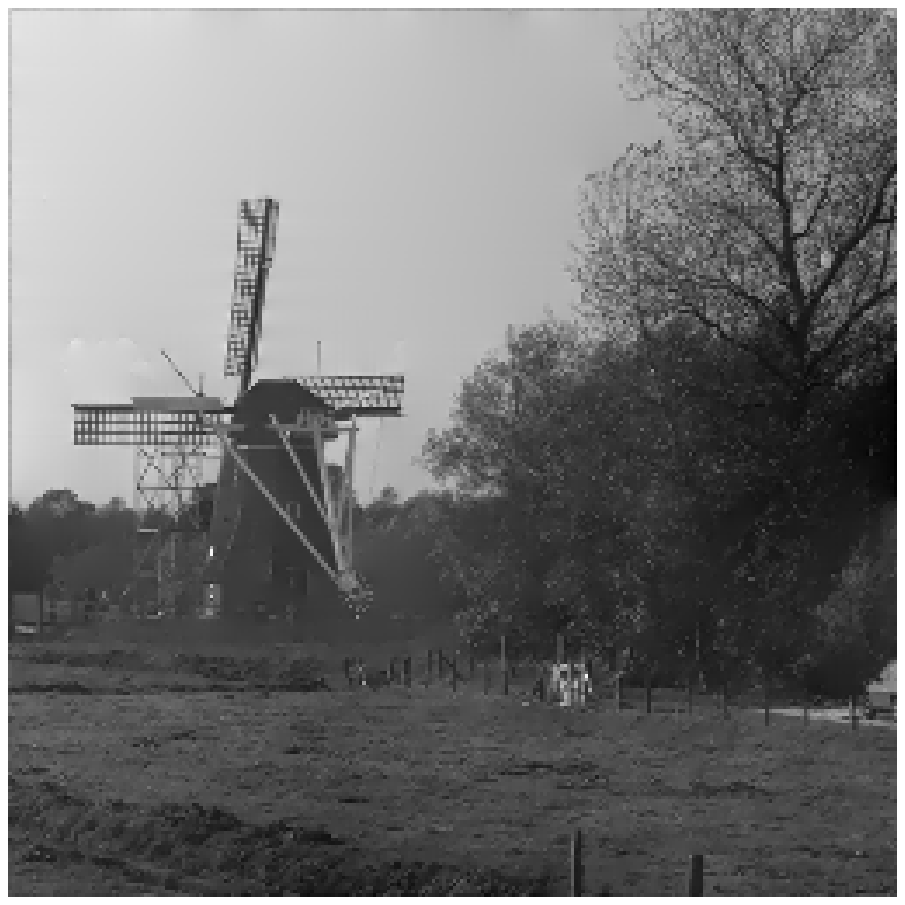}
\end{center}
\caption{First row: (from left to right) 512x512 patches from Hans
van Hateren's images imk01964.imc, imk04089.imc and imk03322.imc. Second
row: their most singular manifolds, obtained as in \cite{Singularities}.
Third row: Their reconstructions. Note that the performance of the
reconstruction is strongly determined by the quality of the
edge-detection}
\label{fig:vH}
\end{figure}

\indent
This is quite reasonable, because then $|\hat{c}(\vec{f})|=\frac{1}{f }
A(\vec{f})$, where $A(\vec{f})=|\hat{\vec{v}}_0(\vec{f}) \cdot \vec{f}|/f$
has a weak dependence on $f$ and varies from one image to another, which
could explain the small exponent $\eta$ in eq.~(\ref{power}).  In Figure
\ref{fig:Lena_kernel} the field $c(\vec{x})$ obtained taking as
$\manifold_{\infty}$ the set given in Figure \ref{fig:Lena_MSM} is
represented. The performance is really good, although quality is lowered
by the (unknown) filtering this image has. This is reasonable because
filtering damages the natural propagation of light, thus the
reconstruction will work up on non-processed natural images. For this we
repeated the process with other, non-filtered images (see Figure
\ref{fig:vH} for one example).  The general performance is good, although
if one border is lost (at the time of edge detection, see
\cite{Singularities}) so is all the structure associated to it, which
seems quite reasonable. 

\indent
Some remarks should be made about the guess for the kernel
$\hat{\vec{g}}(\vec{f})$ given by eq.~(\ref{guess}). First, it is not so
surprisingly simple. It is reasonable thinking about a tridimensional, 
stationary field of light intensity $I(x,y,z)$ propagating out the
secundary sources of light that the objects represent. This scalar field 
sould hence verify $\Delta I\equiv (\frac{\partial^2}{\partial
x^2}+\frac{\partial^2}{\partial y^2}+\frac{\partial^2}{\partial z^2})
I\, =\, 0$ outside those objects. Under apropriate conditions, the
projection of this field on a distant screen could verify that
$\frac{\partial^2 I}{\partial z^2}\approx 0$ at every point in the screen
except those representing the MSM projection. This would imply that the
two-dimensional laplacian of the restriction of $I$ to the screen should
vanish except at the MSM. Then, the image could be reconstructed by
Poissonian diffusion from the MSM, whose propagator is precisely
eq.~(\ref{guess}) for Neumann boundary conditions.

\indent
Second, using the kernel of eq.~(\ref{guess}) and no matter the image
considered, eq.~(\ref{Fourierkernel}) allows reconstructing the correct
$c(\vec{x})$ provided that the set $\manifold_{\infty}$ is large enough:
Taking $\manifold_{\infty}$ as the whole image, eq.~(\ref{Fourierkernel})
turns out to be a trivial identity. It just seems that natural images
allow taking a rather sparse set $\manifold_{\infty}$: the MSM.

\indent
This last remark could be used to enhance the performance of the
reconstruction, by making a more precise determination of the MSM.
Eq.~(\ref{Fourierkernel}), given the kernel eq.~(\ref{guess}), implies
that

\be
div\left( \nabla c \cdot \delta_{\manifold_{\infty}^{c}}\right) = 0
\label{almostPoisson}
\ee

\begin{figure}[htb]
\begin{center}
\hspace*{0cm}
\epsfxsize=8cm
\epsffile{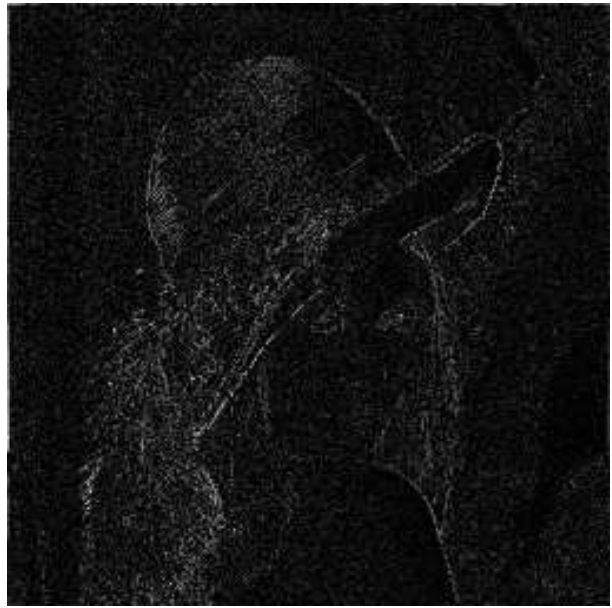}
\end{center}
\caption{Image gotten after applying the laplacian on Lena's image.
The greater is the absolute value, the brighter it is represented. ( A
logarithmic transformation was performed in order to enhance the details )
}
\label{fig:Lena_laplacian}
\end{figure}

\begin{figure}[thb]
\begin{center}
\vbox{
	\hspace*{0cm}
	\epsfxsize=8.5cm
	\epsfysize=5cm
	\epsfbox[50 61 410 292]{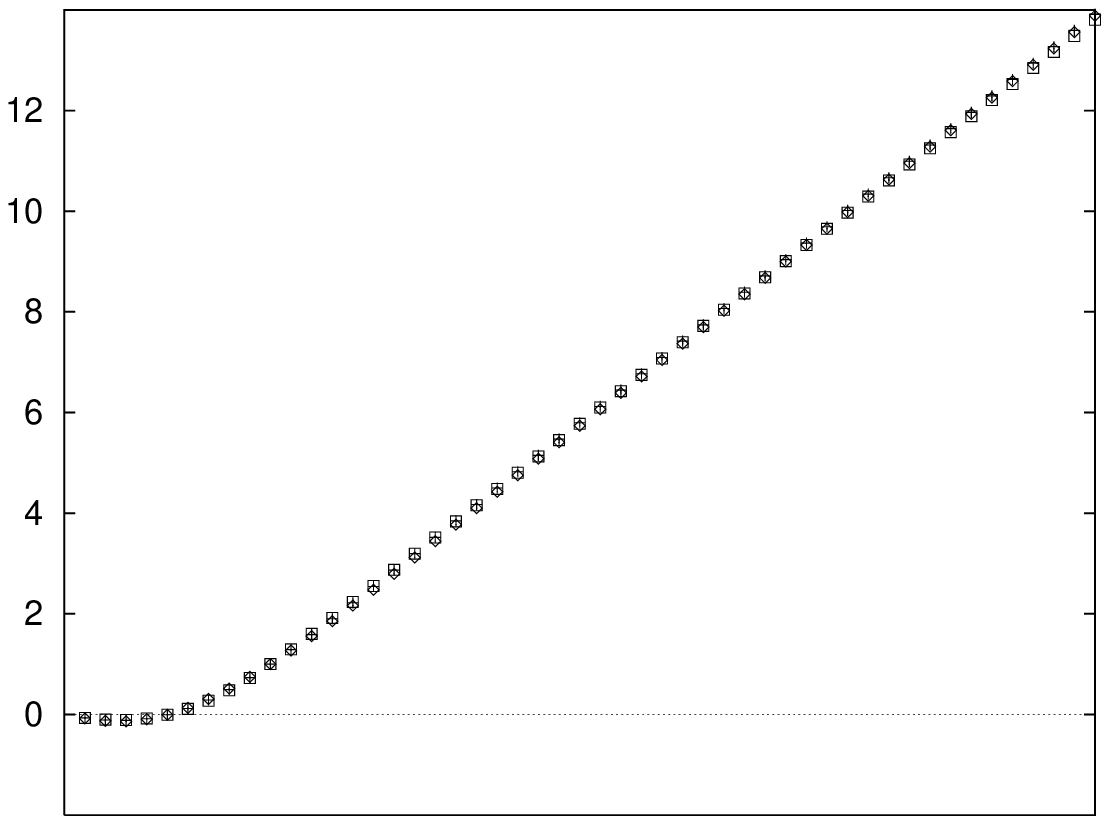}
	\\
	\hspace*{0cm}
	\epsfxsize=8.5cm
	\epsfysize=5cm
	\epsfbox[50 61 410 292]{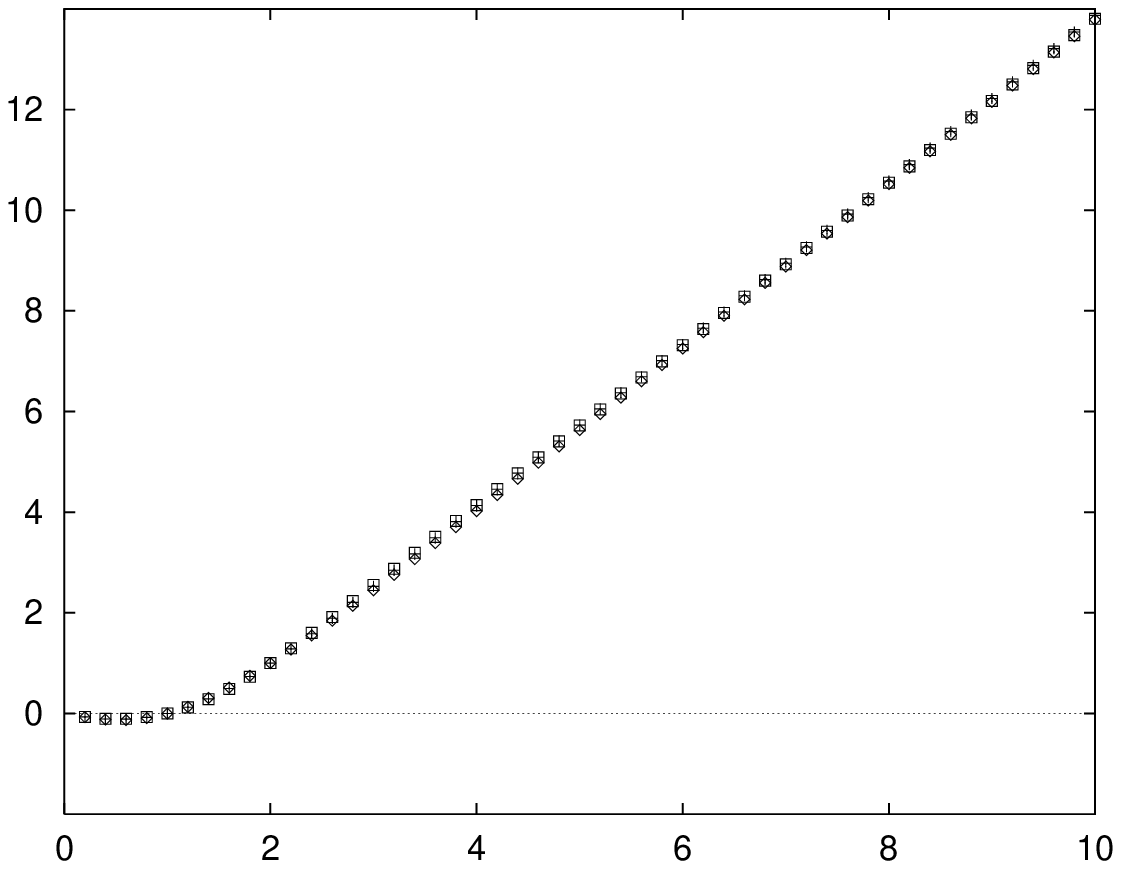}
}
\end{center}
\caption{{\it ESS exponents $\rho(p,2)$ for the original set (OS) 
of images and for the sets obtained after applying a decorrelating filter
(DS) and a Laplacian filter (LS).} They correspond to the horizontal (up)
and the vertical (down) Local Linear Edge Variances. They were calculated
in the same way as in \cite{PRLnuestro}, for the sample formed by van
Hateren's images imk01964.imc and imk04089.imc. The diamonds correspond to
OS, the crosses to DS and the boxes to LS. The graphs almost
perfectly overlap, which indicates that the multifractal structure is
preserved under the transformations (as mentioned in the text). It is also
observed a correspondence in the values between the horizontal and
vertical exponents, from which we conclude that the set is isotropic with
respect to the multifractality.
}
\label{fig:rho}
\end{figure}

\indent
Had the complementary of the MSM ($\manifold_{\infty}^{c}$)  been an open
set, eq.~(\ref{almostPoisson}) would have been a simple Poisson equation
on the contrast, the MSM being the source set. Then, operating the
Laplacian on an image would make it vanish at every point except at those
of the most singular manifold. This operation would allow a fast, direct
way to isolate the MSM and the image would be reduced to a monofractal,
one might say. In Figure \ref{fig:Lena_laplacian} such an operations is
performed over Lena's image, leading to a set rather different than the
MSM previously detected, Figure \ref{fig:Lena_MSM}.  Moreover, the
Laplacian-ed images surprisingly still behave as multifractals, with the
same characteristic parameters. We noticed this repeating the calculations
done in \cite{PRLnuestro} and in \cite{Singularities} over a set of images
got from H. van Hateren (see \cite{Hateren_basis}), and then observing
their scaling properties (see Figure \ref{fig:rho}). To explain this, one
should notice that in fact the MSM is a dense set, so its complementary
set cannot be open and eq.~(\ref{almostPoisson}) does not allow for a
direct splitting. It is also worth noticing that decorrelation of images
(an important strategy in visual coding, see \cite{Atick}) also produces
multifractality of the same kind (see Figure \ref{fig:rho} again). 

\indent
Both transformations (laplacian and decorrelation) intend to be
modifications of the power spectrum, that is, the energy distribution by
frequency of the image (see \cite{Field}). This is equivalent to multiply
the kernel $\hat{g}(\vec{f})$ by $f^2$ or $f$ respectively, and use this
new kernel to propagate from the MSM, according to
eq.~(\ref{Fourierkernel}). We could
even think in a whole family of kernels $\hat{g}_{\alpha}$ which
moduli would scale as $f^{\alpha}$, producing different (in 
principle) multifractals. Those kernels with moduli decaying faster than
$f^{-1}$ generate more saturated images, as the spatial light propagation
would decay slower than with the reconstructor, eq.~(\ref{guess}); those
ones decaying slower in the frequency domain generate images in which
light estinguishes rather close to the source, $\manifold_{\infty}$. But
disregarding the strenght of the propagation, the multifractal hierarchy
remains the same (as partially shown in Figure \ref{fig:rho}). It seems
that the MSM alone explains all the multifractal structure, being this
essentially a geometrical property: in some sense, the singularity
exponent $h$ of each point has only to do with its geometrical situation
with respect to $\manifold_{\infty}$ alone.
	
\indent
Then, the conclusions of our work are:

\begin{enumerate}

\item
Natural images are rather well described by a simple Poissonian diffusion
of the light out of the Most Singular Manifold (MSM), being this a dense
but sparse set in the image.

\item 
In \cite{Singularities}, a conjecture on the relevance from the 
informational point of view of the MSM is issued. Our result shows that 
this could be true: the MSM would allow reconstructing the whole image.
Given that maximization of the information transfer seems to be a very
general learning issue for the sensorial pathways in living beings, 
this suggests that neural structures performing a codification based on
detecting the MSM should be observed (what was first proposed in 
\cite{Singularities}). In fact, eq.~(\ref{guess}) has much to do with 
decorrelation and edge detection, two known features performed by the 
visual neural circuits in mammals.

\item
This coding process seems to be a very efficient one in image processing,
which in turn could be useful in image engeneering. ( for instance, image
compression, light sources modification, etc. )

\item
Multifractality, at least for natural images, is explained by the
only presence of the MSM. In this sense, the
exponent associated to a point measures something like the local
density of edges (segments of the MSM). This issue
could easily be translated to other contexts ( for instance, to fully
developed turbulence ).

\end{enumerate}

\begin{center}
{\bf Acknowledgements}
\end{center}

We are grateful to N\'estor Parga for his support, and to \'Angel Nevado
for fruitful discussions. We want also to acknowledge J.H. van Hateren for
his useful comments and criticism of the manuscript. We also thank Patrick
Meessen for his hintfull corrections. Antonio Turiel is supported by a FPI
grant from the Comunidad Aut\'onoma de Madrid, SPAIN.

\clearpage

\clearpage

\clearpage

\end{document}